\documentclass[10pt,preprintnumbers,amsmath,amssymb]{revtex4}
\usepackage{units}
\usepackage{bbold,euler,newcent}
\usepackage{graphicx}
\usepackage{dcolumn}
\usepackage{bm,bbm,mathtools,esvect}
\usepackage{slashed}

\begin{document}

\title{\LARGE\sf Square-well potential in quaternic quantum mechanics\\ \vspace{5mm}}

\author{\sc SERGIO GIARDINO} 
\email{sergio.giardino@ufrgs.br}
\affiliation{\vspace{3mm} Departamento de Matem\'atica Pura e Aplicada, Universidade Federal do Rio Grande do Sul (UFRGS)\\
Avenida Bento Gon\c calves 9500, Caixa Postal 15080, 91501-970  Porto Alegre, RS, Brazil}

\begin{abstract}
\noindent 
\underline{\sf Abstract}: The one-dimensional infinite square well is the simplest solution of quantum mechanics, and consequently one of the most important. In this article, we provide this solution using the real Hilbert space approach to quaternic quantum mechanics ($\mathbbm{H}$QM).  We further provide the one-dimensional finite as well and a method to generate quaternic solutions from non-degenerate complex solutions.
\end{abstract}

\maketitle
\tableofcontents
\section{\;\sf Introduction\label{I}}
In a few words, quaternion quantum mechanics ($\mathbbm{H}$QM) formulates quantum mechanics using wave functions evaluated
over the quaternic number field ($\mathbbm{H}$). In other words, quaternic wave functions replace the 
usual complex wave functions in Schr\"odinger equation. We will not introduce quaternions in this article, and
a beautiful introduction to quaternions is provided in \cite{Ward:1997qcn}, but we remark that
quaternions are hyper-complex numbers comprising three anti-commutative imaginary units, $i,\,j$ and $k$, where, by way of example, $ij=-ji$. Thus, if  $q\in\mathbbm{H}$, then
\begin{equation}\label{i1}
 q=x_0 + x_1 i + x_2 j + x_3 k, \qquad\mbox{where}\qquad x_0,\,x_1,\,x_2,\,x_3\in\mathbbm{R},\qquad i^2=j^2=k^2=-1.
\end{equation}
 The anti-commutativity of the imaginary units
turns quaternions into non-commutative hyper-complex numbers. We adopt quaternions in symplectic notation, where the definition
(\ref{i1}) is
\begin{equation}\label{i2}
q=z_0+z_1j,\qquad\qquad z_0=x_0+x_1i\qquad\textrm{and}\qquad z_1=x_2+x_3i.
\end{equation}
If quaternions generalize the complex numbers, one can hypothesize that replacing a complex wave function with a quaternic wave function we can generalize quantum mechanics.
The first proposal to $\mathbbm{H}$QM demanded anti-hermitian Hamiltonian operators, and a book by Stephen Adler \cite{Adler:1995qqm} summarizes a large amount of results concerning this approach. 
This anti-hermitian quatenionic theory has several drawbacks, such as the ill-defined classical limit, something that
is clearly described in Adlers's book. Furthermore, anti-hermitian $\mathbbm{H}$QM solutions are rare, complicated, and difficult to interpret. By way of example, we quote several results of this sort
\cite{Davies:1989zza,Davies:1992oqq,Ducati:2001qo,Nishi:2002qd,DeLeo:2005bs,Madureira:2006qps,Ducati:2007wp,Davies:1990pm,DeLeo:2013xfa,DeLeo:2015hza,Giardino:2015iia,Sobhani:2016qdp,
Procopio:2016qqq,Sobhani:2017yee,Hassanabadi:2017wrt,Hassanabadi:2017jiz,Sobhani:2017nfa,Bolokhov:2017ndw,Cahay:2019bqp,Cahay:2019pse,DeLeo:2019bcw,Muraleetharan:2014qma,Sabadini:2017qma}.
We also point out the existence of quaternic applications  in quantum mechanics that cannot be considered $\mathbbm{H}$QM, such as \cite{Arbab:2010kr,Kober:2015bkv,Brody:2011mg,Tabeu:2019cqw,Chanyal:2017rqw,Chanyal:2017xqf}. 

More recently, a novel approach to $\mathbbm{H}$QM eliminated the anti-hermiticity requirement for the Hamiltonian \cite{Giardino:2018lem,Giardino:2018rhs}. Using this approach, it was possible to obtain well defined classical limits \cite{Giardino:2018lem} and a proof of the Virial theorem \cite{Giardino:2019xwm}.  We can confidently state that the real Hilbert space $\mathbbm{H}$QM is physically consistent in a higher level than the anti-Hermitian $\mathbbm{H}$QM. Explicit solutions using this approach are the Aharonov-Bohm effect \cite{Giardino:2016abe}, the free particle \cite{Giardino:2017yke,Giardino:2017nqs}, the quantum Lorentz law \cite{Giardino:2019xwm} and the scattering \cite{Hasan:2020ekd}. 

In this article, we obtained further explicit solutions of $\mathbbm{H}$QM using the real Hilbert space approach: the infinite square well and the finite square well. Although these solutions are probably the simplest quantum systems, they were never solved in $\mathbbm{H}$QM. The novel solutions demanded changes in the topology of domain of the wave function, in the boundary conditions and also the expansion of the quaternic wave function
into Fourier series. Using these changes as a strategy, quaternic solutions were found. The research of simple solutions are  of utmost importance to $\mathbbm{H}$QM. First of all, because they permit a basis where complex quantum mechanics ($\mathbbm{C}$QM) and $\mathbbm{H}$QM can be compared. Additionally, they  will serve as models for more sophisticated quaternic systems that will be build in the future.

The article is organized as follows. In Section \ref{U} we revisit the complex result of the infinite square well to obtain the quaternic
solution. In Secton \ref{F} we apply the procedure to the finite square well. Section \ref{C} rounds off  the article with our conclusions
and future directions.

\section{\;\sf Infinite square well\label{U}}
The quaternic Schr\"odinger equation  is
\begin{equation}\label{u0}
\hbar\frac{\partial\Psi}{\partial t}i=\left(-\frac{\hbar^2}{2m}\nabla^2+V\right)\Psi,
\end{equation}
where the wave function $\Psi$ and the potential $V$ are both quaternic. The imaginary unit $\,i\,$ necessarily multiplies
the right hand side of the time derivative of the wave function, and the equation holds for complex wave functions as well. We used this equation to study
the infinite one-dimensional square well, a problem that is still much studied in quantum mechanics \cite{Belloni:2014isq}, despite its simplicity and its antiquity.
However, the original formulation is too tight to our aim, and a small modification in the potential is necessary. Accordingly, the potential of an infinitely deep square well of width $\,\ell\,$ is 
\begin{equation}\label{u1}
V(x)=\left\{
\begin{array}{ll}
0      & \qquad if\qquad x\in[0,\,\ell] \\
\infty & \qquad if \qquad x\in(-\infty,\,0)\cup (\ell,\,\infty).
\end{array}
\right.
\end{equation}
In its original formulation, the infinite square well potential is defined for the open
interval $\,(0,\,\ell),\,$ but the potential (\ref{u1}) has $\,x=0\,$ and $\,x=\ell\,$ as additional points inside the zero potential region. This  topological change replaces the usual open domain of the wave function with a closed domain that will
enable the existence of  additional solutions. Let us then consider each case.
\subsection{\sf\; complex solution}
In order to have a reference to the quaternic case, we solve the infinite potential well (\ref{u1}) for a complex wave function. Separating the variables, the wave function reads
\begin{equation}\label{u2}
\Psi(x,\,t)= \Phi(x)e^{-i\frac{E}{\hslash}t},
\end{equation}
where $\,E\,$ is the total energy of the system. The time-independent Schr\"odinger equation is simply,
\begin{equation}\label{u3}
\frac{d^2\Phi}{dx^2} =-k^2\Phi \qquad\textrm{with}\qquad k=\frac{\sqrt{2mE}}{\hslash}.
\end{equation}
The general solution for (\ref{u3}) with integration constants $A$ and $B$ is
\begin{equation}\label{u4}
\Phi(x)=Ae^{ikx}+Be^{-ikx}\qquad\mbox{where}\qquad A,\,B\in\mathbbm{C}.
\end{equation}
We propose three different boundary conditions, namely
\begin{equation}\label{u5}
\mbox{i.}\;\;\Phi(0)=\Phi(\ell)=0\qquad\qquad \mbox{ii.}\;\;\Phi(0)=\Phi(\ell)\neq 0\qquad\qquad \mbox{iii.}\;\;\Phi(0)=-\Phi(\ell).
\end{equation}
Conditions (\ref{u5}-ii) and (\ref{u5}-iii) comprise the symmetric and anti-symmetric boundary conditions, and (\ref{u5}-i) is the particular case of (\ref{u5}-ii) that generates the usual solution of the quantum infinite square well. In fact,
condition (\ref{u5}-i) imposes $\,B=-A\,$ in (\ref{u4}) and generates the usual square well solutions set, namely
\begin{equation}\label{u6}
\Phi_n(x)=\sqrt{\frac{2}{\ell}\,}\sin\left(\frac{n\pi}{\ell}x\right)\qquad\qquad E_n=\frac{1}{2m}\left(\frac{n\pi\hslash}{\ell}\right)^2
\qquad\textrm{and}\qquad n\in \mathbbm{N}.
\end{equation}
Condition (\ref{u5}-ii) demands that $\,B=e^{ik\ell}A\,$ and the energy condition will be
\begin{equation}\label{u7}
k\ell=2n\pi+\varphi_0,\qquad\textrm{where}\qquad\varphi_0\in[0,\,2\pi]\qquad\textrm{and}\qquad n\in\mathbbm{Z}.
\end{equation}
Demanding orthogonality between the physical states, we obtain that $\,\varphi_0=0,\,$ and thus the solutions set is
\begin{equation}\label{u8}
\Phi_n(x)=\sqrt{\frac{2}{\ell}\,}\cos\left(\frac{2n\pi}{\ell}x\right)\qquad\qquad E_n=\frac{1}{2m}\left(\frac{2n\pi\hslash}{\ell}\right)^2
\qquad\textrm{and}\qquad n\in\mathbbm{N}.
\end{equation}
By the same token, (\ref{u5}-iii) is
\begin{equation}\label{u9}
\Phi_n(x)=\sqrt{\frac{2}{\ell}\,}\cos\left[\frac{(2n+1)\pi}{\ell}x\right]\qquad\qquad E_n=\frac{1}{2m}\left[\frac{(2n+1)\pi\hslash}{\ell}\right]^2
\qquad\textrm{and}\qquad n\in\mathbbm{Z}^+.
\end{equation}
The single physical change that comes from conditions (\ref{u5}-ii) and (\ref{u5}-iii) comprises the energy gaps between the physical states, respectively
\begin{equation}\label{u010}
\mbox{i.}\;\;\frac{E_{n+1}-E_n}{\frac{1}{m}\left(\frac{\pi\hslash}{\ell}\right)^2}= n+\frac{1}{2} 
\quad\qquad\mbox{ii.}\;\; \frac{E_{n+1}-E_n}{\frac{1}{m}\left(\frac{\pi\hslash}{\ell}\right)^2}= 4\left(n+\frac{1}{2}\right)
\quad\qquad\mbox{iii.}\;\; \frac{E_{n+1}-E_n}{\frac{1}{m}\left(\frac{\pi\hslash}{\ell}\right)^2}= 4\left(n+\frac{1}{2}\right)+2.
\end{equation}
We observe that the behavior of the energy gaps are different in each case, that are physically nonequivalent and hence correspond to novel solutions.
The physical observables that remain preserved irrespective of the case are
\begin{equation}\label{u10}
\langle x\rangle=\frac{\ell}{2}\qquad \langle p\rangle=0\qquad\langle p^2\rangle=2mE.
\end{equation}
The observables $\langle x^n\rangle$ may have slight differences, but the uncertainty relation $\Delta x\Delta p\geq \hslash/2$ remains
untouched. Thus, the wave functions are not identical probability densities, and we may expect deviations of the expectation values from the usual case, although they have identical results for position and momentum. 

These results may be considered an interesting pedagogical exercise for quantum mechanics classes. However, the new potential (\ref{u1}) and piecewise continuous wave functions are necessary conditions for $\mathbbm{H}$QM, as follows in the next case.
\subsection{\;\sf quaternic solution}
In this case, the time independent solutions for (\ref{u5}) will be sought as
\begin{equation}\label{u11}
\Phi(x)=a\Big(\cos kx e^{i\phi_0}+\sin kx e^{i\xi_0}j\,\Big),\qquad\textrm{with}\qquad a\in\mathbbm{R}\qquad\textrm{and}\qquad
\phi_0,\,\xi_0\in[0,\,2\pi].
\end{equation}
We notice that the normalization constant is  real because the Hilbert space is real, in agreement with  
\cite{Giardino:2018lem,Giardino:2018rhs}. It is important to notice that (\ref{u11}) does not represent a wave superposition, because the quaternic and complex components are orthogonal. Such a wave function was firstly proposed in \cite{Giardino:2018rhs} as the basis element of the the quaternic Fourier series. Thus, we have an interesting
analogy between the complex case and the quaternic case. The boundary condition (\ref{u5}-i) is too tight because it forces $\Phi(x)=0$ everywere. On the other hand,
imposing  (\ref{u5}-ii) and (\ref{u5}-iii) we respectively obtain
\begin{equation}\label{u12}
k\ell=2n\pi\qquad\textrm{and}\qquad k\ell=(2n+1)\pi.
\end{equation}
The normalization constant is immediately obtained to be
\begin{equation}
a=\frac{1}{\sqrt{\ell\,}\,}.
\end{equation}
The stationary state elements comprise a  basis elements that satisfies the orthogonality condition \cite{Giardino:2018lem,Giardino:2018rhs}
\begin{equation}
\int_0^\ell \Psi_k\overline{\Psi}_{k'}\,dx\,=\,\delta_{kk'},
\end{equation}
where we used
\begin{equation}
 \Psi_k\overline\Psi_{k'}\,=\,\frac{1}{\ell}\left[\,\cos(k-k')x\,+\,\sin(k-k')x\,e^{i(\phi_0+\xi_0)}\,j\,\right].
\end{equation}
The expectation value definition for a real Hilbert space \cite{Giardino:2018lem,Giardino:2018rhs}, namely
\begin{equation}\label{u13}
\langle\mathcal O\rangle= \frac{1}{2}\int dx^3\Big[\big(\mathcal{O}\Psi\big)\overline\Psi^{\,} +\Psi\big(\overline{\mathcal{O}\Psi}\big) \Big],
\end{equation}
enables us to recover the expectation values of (\ref{u8}-\ref{u9}). 
On the other hand, the quaternic square well is more restrictive than the complex square well, because
the condition $k\ell=n\pi$ that generates  (\ref{u6}) produces a trivial solution in $\mathbbm{H}$QM. 
This seems counter-intuitive, because it was expected that additional degrees of freedom would 
allow additional solutions, and not fewer solutions. One could hypothesize that this is an effect of the non-commutativity of the wave function. However, quaternic wave functions have additional boundary conditions concerning the purely quaternic part of the wave function, and these additional conditions eliminated the solutions whose energy is identical to (\ref{u6}).

\subsection{\;\sf combined quaternic solution\label{aqs}}
In this section we study quaternic solutions of the form
\begin{equation}\label{u14}
\Psi_{np}\,=\,A_n\,\psi_n+B_p\,\psi_p \,	j\qquad{\rm where}\qquad A_n,\,B_p\in\mathbbm{C},\qquad n,\,p\in\mathbbm{N}
\end{equation}
and $\psi_n$ are complex wave functions that satisfy (\ref{u0}-\ref{u1}). Thus, (\ref{u14}) comprises two non-degenerate 
solutions of $\mathbbm{C}$QM, where $\,n\neq p,\,$
 jointed in a single quaternic structure. We name such functions the combined quaternic solution.
 A naive orthogonality condition for such kind of wave function is something like
\begin{equation}\label{u15}
\big\langle\,\Psi_{np},\,\Psi_{n'p'}\big\rangle\,=\,a_n\,\delta_{nn'}+b_p\,\delta_{pp'}
\end{equation}
where $a_n$ and $b_p$ are real constants. This condition is not sufficiently tight to determine the coefficients of a quaternic function
through a Fourier expansion. An additional parallelism condition will be necessary, as we shall see in the example below. We remember
that we can define an angle between two quaternions using the inner product (\ref{u13}), and thus we obtain the angle between $\Psi$ and $\mathcal O\Psi$ after dividing (\ref{u13}) by the norm of each quaternion. The concepts of orthogonality and parallelism are immediate from this definition \cite{Ward:1997qcn}. Let us then consider the quaternic wave function 
\begin{equation}\label{u16}
\Psi_{np}=\sqrt{\,\frac{2}{\ell}\,}\,\Bigg[\,\cos\theta_{np}\,\sin\left(\frac{n\pi}{\ell}x\right)e^{-i\frac{E_n}{\hbar}t}+\sin\theta_{np}\,\sin\left(\frac{p\pi}{\ell}x\right)e^{i\frac{E_p}{\hbar}t}j\,\Bigg],
\end{equation}
which satisfies the boundary condition (\ref{u5}-i) and whose energies come from (\ref{u6}). Using the energy operator $\,\widehat E=\big(\hbar\partial_t|i\big),\,$  where $\,(a|b)f\,=\,a\,f\,b,\,$ we get
\begin{equation}\label{u17}
\big\langle E_{np}\big\rangle\,=\,E_n\cos^2\theta_{np}\,+\, E_p\sin^2\theta_{np},
\end{equation}
and the complex case is recovered either for $n=p$  or in the limit $\theta_{np}=0$.
The most general solution in a real Hilbert space is 	
\begin{equation}\label{u18}
\Psi(x,\,t)=\sum_{n,p\in\mathbbm{N}}c_{np}\Psi_{np}(x,\,t)\qquad{\rm where}\qquad c_{np}\in\mathbbm{R}.
\end{equation}
Imposing the constraint 
\begin{equation}
p=p(n),\qquad\mbox{where}\qquad p:\mathbbm{N}\to\mathbbm{N}
\end{equation}
where $p(n)$ have to be injective. Consequently $\,\delta_{nn'}=\delta_{pp'}\,$ and we are allowed to make $\,\theta_{np}\to\theta_n\,$ to obtain the orthogonality condition,
\begin{eqnarray}\label{u19}
\nonumber
\big\langle\Psi_{np},\,\Psi_{n'p'}\big\rangle&=&\cos\theta_n\cos\theta_{n'}\delta_{nn'}+\sin\theta_n\sin\theta_{n'}\delta_{pp'}\\
&=&\delta_{nn'}.
\end{eqnarray}
We point out that the orthogonality condition requires that $\theta_n=\theta_{n'}$, the quaternic 
parallelism condition.
We remember that  each basis element of a complex Fourier series is a complex exponential defined in a bi-dimensional plane, and thus each basis element has a single orthogonal direction direction associated to it. However, each quaternic basis element is defined inside a
four-dimensional hypercomplex space, and consequently has three orthogonal directions associated to it. Remembering that two quaternions $p$ and $q$ are parallel if $p\bar q$ is real \cite{Ward:1997qcn}, 
from (\ref{u16}) and (\ref{u19}) we observe that the parallel direction requires a fixed value to $\theta_{np}$.
Consequently, the orthogonality condition selects the direction parallel to $\Psi_{np}$
in the arbitrary function $\Psi$, and using (\ref{u18}-\ref{u19}) we finally obtain
\begin{equation}\label{u21}
\big\langle\,\Psi(x,\,0),\,\Psi_{np}(x,\,0)\,\big\rangle= c_{np}=c_n.
\end{equation}
Therefore, the wave function (\ref{u16}) defines a subspace of the real Hilbert space where each basis element is an independent physical state, in total agreement with $\mathbbm{C}$QM. In the simplest case,
where $\theta_{np}=\theta$, the minimum difference between physical states is
\begin{equation}
\frac{ E_{(n+1)p}-E_{np}}{\frac{1}{m}\left(\frac{\hbar\pi}{\ell}\right)^2}=\left(n+\frac{1}{2}\right)\cos^2\theta.
\end{equation}
As expected, this result recovers the complex case (\ref{u010}-i) for $\theta=0$, which means that the quaternion case is parallel to the complex case. The theta parameter thus quantify the deviation from the parallelism between the pure complex and the quaternic case. Physically, the greater the deviation, the smaller the gap between the physical states. However the precise physical meaning of the $\theta$ parameter is an open question for future research.

\section{\;\sf Finite square well\label{F}}
In this case, we adopt the usual real potential
\begin{equation}\label{f1}
V(x)=\left\{
\begin{array}{ll}
0&\qquad\textrm{if}\qquad |x|\leq\ell/2\\
V_0&\qquad \textrm{if}\qquad |x|> \ell/2
\end{array}
\right.\qquad\textrm{where}\qquad V_0>0.
\end{equation}
Differently from (\ref{u1}), this the potential (\ref{f1}) is identical to the usual complex case. In the next section, we 
recover the usual results that can be found in every textbook of quantum mechanics, and that will be the reference to the quaternic cases.
\subsection{\sf the complex solution}
The finite well admits bound solutions and scattered solutions as well. In both of them, the complex wave function and its first derivative are continuous at the boundaries of the well, where $x=\pm\ell/2$. In the case of bound states, where $E<V_0$, the time independent $\mathbbm{C}$QM wave function is
\begin{equation}\label{f2}
\phi(x)=\left\{
\begin{array}{l}
A e^{\kappa x}\qquad\quad\; {\rm  for}\qquad\, x<-\ell/2 \\
B\cos kx\qquad {\rm  for}\qquad |x|<\ell/2 \\
A e^{-\kappa x} \qquad\;\; {\rm  for}\qquad \,x>\ell/2 
\end{array}
\right.
\end{equation}
where $k$ and $\kappa$ are
\begin{equation}\label{f3}
\kappa=\frac{\sqrt{\,2m\,|V_0-E|\,}}{\hslash}\qquad\qquad\mbox{and}\qquad\qquad k=\frac{\sqrt{2mE\,}}{\hslash}.
\end{equation}
and the normalization constants are
\begin{equation}\label{f4}
A^2=\frac{ke^{\kappa\ell}}{\frac{k}{\kappa}+\frac{1}{2}\left(1+\frac{\kappa^2}{k^2}\right)\Big(k\ell+\sin k\ell\Big)}\qquad\qquad
B^2=\frac{k\left(1+\frac{\kappa^2}{k^2}\right)}{\frac{k}{\kappa}+\frac{1}{2}\left(1+\frac{\kappa^2}{k^2}\right)\Big(k\ell+\sin k\ell\Big)}.
\end{equation}
The quantization of the finite square well is numerically obtained from the transcendental equation
\begin{equation}\label{f5}
\sqrt{\frac{V_0}{E}-1\,}=\tan\sqrt{\frac{m}{2}(V_0-E)\,}\frac{\ell}{\hbar}.
\end{equation}
On the other hand, the scattered states have $V_0<E$ and the wave function is
\begin{equation}\label{f6}
\varphi(x)=\left\{
\begin{array}{l}
e^{i\kappa x}+Re^{-i\kappa x}\qquad\quad\quad\; {\rm  if}\qquad\, x<-\ell/2 \\
C\cos kx+D\sin kx\qquad {\rm  if}\qquad |x|<\ell/2 \\
T\, e^{i\kappa x} \qquad\qquad\qquad\quad\;\;\, {\rm  if}\qquad \,x>\ell/2, 
\end{array}
\right.
\end{equation}
where $C,\,D,\,R\,$ and $\,T\,$ are complex constants. The continuity of the wave function and its first derivative at the wedges of the well permit us obtain
\begin{equation}\label{f7}
 R=i\frac{\sin k\ell}{2}\frac{k^2-\kappa^2}{k\kappa} T\qquad{\rm and}\qquad 
 T=2 e^{-i\kappa\ell}\frac{2\cos k\ell+i\frac{k^2+\kappa^2}{k\kappa}\sin k\ell}{4\cos^2k\ell+\left(\frac{k^2+\kappa^2}{k\kappa}\right)^2\sin^2k\ell}.
\end{equation}
$|R|^2$ and $|T|^2$ are respectively the coefficients of reflection and transmission, which satisty $|R|^2+|T|^2=1$. Let us keep the above well known results as a reference to the quaternic solutions that follow.
\subsection{\;\sf quaternic bound state\label{Fb}}
We expect bound states inside the well for energies within the range $(0,\,V_0).$ Therefore
we propose the solution
\begin{equation}\label{f8}
\Phi(x)=\left\{
\begin{array}{ll}
A\, e^{\kappa x}&\qquad\textrm{if}\qquad\; x<-\ell/2\\
B\Big(\cos kx \,e^{i\phi_2}+\sin kx \,e^{i\xi_2}j\Big)&\qquad\textrm{if}\qquad |x|<\ell/2\\
C\, e^{-\kappa x}&\qquad\textrm{if}\qquad\; x>\ell/2,
\end{array}
\right.
\end{equation}
where $A,\,B$ and $\,C\,$ are real integration constants. Differently from the complex case, we impose  the continuity 
for the absolute value of the wave function and its first derivative at $\,x=\pm\ell/2.\,$ The continuity of the absolute value of the quaternic wave functions ($|\Phi|^2$) gives
\begin{equation}\label{f9}
A^2=C^2=B^2e^{\kappa\ell}
\end{equation}
The normalization also leads to
\begin{equation}\label{f10}
B^2=\frac{\kappa}{1+\kappa\ell}
\end{equation}
On the other hand, using (\ref{f4}) and the continuity of the absolute value the wave function's first derivative ($|\Phi'|^2$) imposes
\begin{equation}\label{f11}
k^2=\kappa^2\qquad\qquad\mbox{and consequently}\qquad\qquad E=\frac{V_0}{2}.
\end{equation} 
Therefore the energy is not quantized. However, the continuity of the absolute value of the wave function additionally impose a symmetry condition and the consequent quantization where
\begin{equation}\label{f12}
\Phi\left(\frac{\ell}{2}\right)=\Phi\left(-\frac{\ell}{2}\right),\qquad\qquad m_n=\frac{4}{V_0}\left[\frac{n\pi\hbar}{\ell}\right]^2\qquad\textrm{and}\qquad n\in\mathbbm{N}.
\end{equation}
Conversely, the anti-symmetric case comprises
\begin{equation}\label{f13}
\Phi\left(\frac{\ell}{2}\right)=-\Phi\left(-\frac{\ell}{2}\right),\qquad\qquad m_n=\frac{4}{V_0}\left[\frac{(2n-1)\pi\hbar}{\ell}\right]^2\qquad\mbox{and}\qquad n\in\mathbbm{N}.
\end{equation}
Remembering that the potential $\,V_0\,$ is a free parameter, we conclude that the mass $\,m_k\,$ is quantized, and the existence of a bound state is determined by the existence of a  particle whose mass fits the conditions (\ref{f7}-\ref{f8}).
In the limit of infinite energy, where $\,\kappa\to\infty,\,$ the normalization integrals vanish outside the well and we obtain a massless single state of divergent energy trapped inside an infinite square well. These features make the quaternic solution new and  without a complex counterpart. Imposing similar boundary conditions in $\mathbbm{C}$QM, condition (\ref{f11}) is replaced with
\begin{equation}\label{f013}
1-\frac{V_0}{E}=\cot^2\frac{k\ell}{2},
\end{equation}
and infinite numerically defined energy levels come. Consequently, the quaternic solution is something completely novel.
\subsection{\;\sf scattering state\label{Fs}}
The time independent solutions for $V_0<E$ are
\begin{equation}\label{f14}
\Phi(x)=\left\{
\begin{array}{ll}
e^{i\kappa x}+R\,e^{-i\kappa x}&\qquad\textrm{if}\qquad x<-\ell/2\\
T\Big(\cos kx \,e^{i\phi_2}+\sin kx \,e^{i\xi_2}j\Big)&\qquad\textrm{if}\qquad |x|<\ell/2\\
D\,e^{i\kappa x}&\qquad\textrm{if}\qquad x>\ell/2,
\end{array}
\right.
\end{equation}
and the integration constants $R,\,T\,$ and $\,D\,$ are real.
The the transmitted waves are related according to 
\begin{equation}\label{f16}
R^2+T^2=1.
\end{equation}
We interpret (\ref{f16}) as a constraint that selects the physical states. Imposing the continuity of the absolute value of the wave function at $\,x=\pm\,\ell/2\,$, we obtain
\begin{equation}\label{f17}
1+R^2+2R\cos \kappa\ell\,=\,T^2\,=\,D^2.
\end{equation}
However, coherently with the asymmetrical character of the problem, we
impose the continuity of $\,|\Phi'|\,$  at the scattering point $\,x=\ell/2\,$ only, so that
\begin{equation}\label{f18}
1+R^2-2R\cos \kappa\ell\,=\,T^2\left(\frac{k}{\kappa}\right)^2.
\end{equation}
We stress that this boundary condition is totally novel. From (\ref{f16}-\ref{f17}) we get
\begin{equation}\label{f19}
R=-\cos \kappa\ell\qquad{\rm and}\qquad T^2=\sin^2 \kappa\ell.
\end{equation}
Consequently the quantization condition is
\begin{equation}\label{f21}
\tan^2\kappa\ell=4\left(\frac{E}{V_0}-1\right)
\end{equation}
and the reflection and transmission coefficients are such that
\begin{equation}\label{f20}
R^2=\frac{\frac{1}{4}\frac{V_0}{E}}{1-\frac{3}{4}\frac{V_0}{E}},\qquad\mbox{and}\qquad T^2=\frac{1-\frac{V_0}{E}}{1-\frac{3}{4}\frac{V_0}{E}}.
\end{equation}
Therefore, the transmission increases with the energy and asymptotically approximates the total transmission, as expected. On the other hand, there are two points of 
intersection around $p\ell=(2n+1)\pi/2$. Thus the allowed states are quantized, as expected. However, we point out that the quantization condition is similar to (\ref{f013}), that quantizes complex bound states. This similarity  between the complex bound states and the quaternic scattering states is a novel and interesting result. Furthermore, the quaternic solution is again more restrictive than the complex scattering, where the energy is not quantized. The differences between the complex and quaternic cases are consequently remarkable.
Finally, we have two possible limit cases whose  lowest approximations give
\begin{eqnarray}\label{f22}
&&V_0\approx E\;\qquad\Rightarrow T^2\approx 1\qquad{\rm and}\qquad \kappa \pi=2n\pi\\
&&V_0\ll E\qquad\Rightarrow R^2\approx 1\qquad{\rm and}\qquad \kappa \pi=(2n+1)\pi \nonumber
\end{eqnarray}
Remembering that $E>V_0$, these limit cases indicate the situations when free particles have quantum energies, which may be obtained adiabatically within the limit $V_0\to 0$. Furthermore, the result indicates that low energy free particles must have quantum energies, something that is not seen in $\mathbbm{C}$QM. Consequently, the finite square well leaves interesting directions for future research.  
\section{\;\sf Conclusion\label{C}}
In this article we presented the quaternic solutions to the simplest one-dimensional quantum problems: the infinite and the finite quantum wells.  Two innovations were decisive to obtain these solutions: the real Hilbert space and the quaternic Fourier series, and one of the most interesting novelties are the combined solutions, where complex solutions of arbitrary energies were used to obtain a single quaternic solution. 

The results show that the quaternic solutions are more restrictive than the complex ones, and this feature is due to the additional boundary conditions imposed to quaternic wave functions. However, the quaternic solutions are physically different from the complex ones, and thus different physical system can be modeled using these techniques. 

In summary, the article presents a new technical framework that enable to built novel quantum solutions. There are many possible directions for  future research, simply because the whole non-relativistic quantum mechanics can be studied using these new implements. Intersting results may be obtained from the previous results in $\mathbbm{C}$QM, for example concerning application of the rectangular potentials to the Stark effect, and also to supersymmetry \cite{Belloni:2014isq}. However, the most simple possibilities shall be obtained building different kinds of combined solutions. Moreover, the relativistic quantum theory is also open to explore using this new machinery and other quaternic alternatives \cite{Chanyal:2017rqw,Chanyal:2017xqf}. Consequently we expect further new solutions in the near future.

%
%
%
%

\bibliographystyle{unsrt} 
\bibliography{bib_poco_potencial}
\end{document}